\documentclass[apj]{emulateapj}

\shorttitle{$^{7}$Be in V5668 Sgr and V2933 Oph}
\shortauthors{Tajitsu et al.}

\newcommand{\iiihe}{$^{3}$He}
\newcommand{\ivhe}{$^{4}$He}

\newcommand{\nai}{Na\,{\sc i}}

\newcommand{\cai}{Ca\,{\sc i}}
\newcommand{\caii}{Ca\,{\sc ii}}

\newcommand{\feii}{Fe\,{\sc ii}}

\newcommand{\hi}{H\,{\sc i}}

\newcommand{\hg}{H$\gamma$}

\newcommand{\kms}{km\,s$^{-1}$}
\newcommand{\vrad}{$v_{\rm rad}$}

\newcommand{\bevii}{$^{7}$Be}
\newcommand{\beviiii}{$^{7}$Be\,{\sc ii}}
\newcommand{\lii}{Li\,{\sc i}}

\newcommand{\livii}{$^{7}$Li}
\newcommand{\liviii}{$^{7}$Li\,{\sc i}}

\begin{document}


\title{The \beviiii\ Resonance Lines in Two Classical Novae 
V5668 Sgr and V2944 Oph}


\author{Akito\ Tajitsu\altaffilmark{1}}
\altaffiltext{1}{Subaru Telescope, National Astronomical Observatory of Japan, 650 North A'ohoku Place, Hilo, HI 96720, USA}
\email{tajitsu@naoj.org}

\author{Kozo\ Sadakane\altaffilmark{2}}
\altaffiltext{2}{Astronomical Institute, Osaka Kyoiku University, Asahigaoka, Kashiwara, Osaka 582-8582, Japan}

\author{Hiroyuki\ Naito\altaffilmark{3}}
\altaffiltext{3}{Nayoro Observatory, 157-1 Nisshin, Nayoro, Hokkaido 096-0066, Japan}

\author{Akira\ Arai\altaffilmark{4}}

\author{Hideyo\ Kawakita\altaffilmark{4}}
\altaffiltext{4}{Koyama Astronomical Observatory, Kyoto Sangyo University, Motoyama, Kamigamo, Kita-ku, Kyoto 603-8555, Japan}

\author{Wako\ Aoki\altaffilmark{5}}
\altaffiltext{5}{National Astronomical Observatory of Japan, 2-21-1 Osawa, Mitaka, Tokyo 181-8588, Japan}



\begin{abstract}
We report spectroscopic observations of the resonance lines of 
 singly ionized \bevii\ in 
the blue-shifted absorption line systems found in the 
post-outburst spectra of two classical novae --
 V5668 Sgr (Nova Sagittarii 2015 No.\,2) and V2944 Oph (Nova Ophiuchi 2015). 
The unstable isotope, \bevii, should has been created during
 the thermonuclear runaway (TNR) of these novae 
 and decays to form \livii\ within a short period (a half-life of 53.22 days). 
Confirmations of \bevii\ are the second and the third ones 
 following the first case found in V339 Del by \cite{2015Natur.518..381T}.
The blue-shifted absorption line systems in both novae 
are clearly divided into two velocity components, both of which contain \bevii.
This means that the absorbing gases in both velocity components consist
 of products of TNR.
We estimate amounts of \bevii\ produced during outbursts of 
 both novae and conclude that significant \livii\ should have been created. 
These findings strongly suggest that the explosive production of \livii\ 
via the reaction \iiihe($\alpha$,$\gamma$)\bevii\ and subsequent decay to \livii\ occurs frequently among classical novae and contributes to the
process of the Galactic Li enrichment.
\end{abstract}





\section{Introduction}
Lithium (Li) is a key element in the study of the chemical evolution of 
the universe because it likely has been produced in various sites and events --
Big Bang nucleosynthesis, 
interactions between energetic cosmic rays and interstellar matter, 
evolved low-mass stars, novae, and supernova explosions.
The observed Li evolutionary curve has a plateau for young Galactic ages 
($<$ 2.5 Gyr) followed by a steep rise.
This indicates that a relatively low-mass stellar component with long 
lifetimes is a major source of Li in the recent universe
(\citealt{1999AaA...352..117R, 2001AaA...374..646R, 2012AaA...542A..67P}). 
Some low-mass evolved stars have indeed been found to have Li-enriched 
surfaces (e.g., \citealt{2005AaA...439..227M}).
However, there has been no observational confirmation that such stars 
supply Li to interstellar medium.
Li contained in stellar surface could easily be depleted by convection,
because Li should be destroyed inside stars where temperature is higher than 
2.5 million K.

Novae, which evolve from low-mass binaries, 
are expected to be
one of the candidates for Li suppliers in the recent universe,
because they experience explosive nucleosynthesis and mass-loss simultaneously.
\cite{1971ApJ...164..111C} made the first theoretical study of Li 
production in novae.
They predicted that the radioactive isotope of beryllium, \bevii, produced
via the reaction \iiihe($\alpha$, $\gamma$)\bevii\ during 
the TNR in the accumulated thin gas layer on the surface of a white dwarf (WD).
The produced \bevii\ will be blown away from the surface of a WD 
 by the outburst wind, then decays to form \livii\ in the cooler interstellar environment within a short time (a half-life of 53.22 days).
In the 1990s, details of this process, which is called as the Cameron-Fowler process, were studied by theoretical analyses (\citealt{1993AaA...279..173B,1996ApJ...465L..27H,1998ApJ...494..680J}).

Recently, we have reported the detection of \bevii\ in the post-outburst UV spectra of the classical nova V339 Del (\citealt{2015Natur.518..381T}).
Because the resonance doublet lines of Be\,{\sc ii} are located at
 $\sim$3130 \AA\  where the telluric absorption mainly caused by ozone 
severely obstructs observations from  ground-based telescopes, 
 quantitative studies of the stellar Be abundances have been enabled 
 in 1990s by the advent of 8-10 m class telescopes on high mountains.
They are equipped with high-resolution spectrographs having high sensitivities
 in the near-UV region [e.g., \citealt{1999AJ....117.1549B} (Keck\,{\sc i} telescope); \citealt{2000AaA...362..666P} (VLT)].
Extensive analyses of stellar Be abundances were conducted for 
 large samples of metal-poor stars (\citealt{2011ApJ...743..140B}) 
 and solar analogs (\citealt{2011PASJ...63..697T}).
In the case of V339 Del, we found that the absorption lines of Be\,{\sc ii} 
 were purely originating from \bevii\ instead of the stable $^{9}$Be.
Therefore,
it was the first observational confirmation of the Cameron-Fowler process in
classical novae.
The \bevii\ was found in highly blue-shifted (radial velocities, \vrad $\sim$ 1000 \kms) absorption lines, which are assumed to correspond to nova ejecta blown off from the surface of a WD by the outburst wind.
Through the direct comparison of absorption strengths between \beviiii\ and \caii,
we estimated that the \bevii\ abundance in the ejecta of V339 Del could be higher than the pre-existing theoretical prediction (\citealt{1998ApJ...494..680J}).
\cite{2015arXiv150608048I} reported the presence of the 
\lii\,$\lambda$6708 line in the blue-shifted absorption line system 
in the early phase (7--13 days after the outburst) spectra of V1369 Cen 
(Nova Centauri 2013).
The overabundance of Li in this nova amounts to the order of $10^{4}$
with respect to the solar photospheric composition.

Now it is quite interesting to know how commonly this \bevii\ (=\livii) 
production occurs among various types of classical novae and evaluate 
their contribution to the Li evolution in the Galaxy.
Prompted by these interests, we have performed high resolution spectroscopic observations
of two Galactic novae discovered in  March 2015 --
V5668 Sgr and V2944 Oph, during their early decline phases.

V5668 Sgr (Nova Sagittarii 2015 No.\,2 = PNV J18365700$-$2855420) was discovered as a bright 6.0 magnitude (unfiltered) source by John Seach on 2015 March 15.634 {\sc UT} and announced in the American Association of Variable Star Observers (AAVSO) Alert Notice (\citealt{2015AAN...512....1W}).
The nova recorded its optical maximum ($V = 4.3$) on March 21.67 {\sc UT} (MJD=57102.67).
Its optical magnitudes stayed close to the maximum ($V \sim 4.5$ -- 6.5) 
for about 80 days,  
then showed a rapid decline by dust formation (\citealt{2015ATel.7299....1B,2015ATel.7303....1B,2015ATel.7862....1G}).
The possible candidates of the progenitor were found in the USNO-B1.0 catalogue 
(\citealt{2003AJ....125..984M}) within 6\arcsec\ from the nova (USNO-B1.0 0610-078494.3, B1=17.17, R1=16.82, R2=16.42, I=15.55mag; USNO-B1.0 0610-0784923, R1=14.36, B2=13.62, R2=13.65, I=15.08 mag), and in
the Galaxy Evolution Explorer ($GALEX$; \citealt{2005ApJ...619L...1M}) catalogue of NUV (1171--2831 \AA) sources 3\arcsec.6 from the nova ($GALEX$ J183656.8-285539, NUV=17.323 mag) (\citealt{2015AAN...512....1W}).
From a spectroscopic observation obtained immediately after the discovery,
\cite{2015ATel.7230....1W} reported that it is an \feii\ type nova according to the classification introduced by \cite{1992AJ....104..725W}.

V2944 Oph (Nova Ophiuchi 2015 = PNV J17291350$-$1846120) was discovered as a 12.2 magnitude (unfiltered) source by Yukio Sakurai on 2015 March 29.766 {\sc UT} (\citealt{2015AAN...516....1W}).
After its discovery, the nova showed a gradual rise to its optical maximum at $V = 9.0$ on April 14.75 {\sc UT} (MJD = 57126.75).
Then, it showed a quick decline to $V \sim$ 11.5 within the following few days, and stayed 
at $V \sim 12$ for $>$ 80 days.
In the USNO-B1.0 catalogue, a possible candidate of the progenitor was found as a $\sim$18 magnitude star at the position of the nova (\citealt{2015AAN...516....1W}).
Its spectrum during the pre-maximum phase showed He/N type characteristics 
(\citealt{2015ATel.7339....1D}).
However,  \cite{2015ATel.7367....1M} concluded that it is an \feii\ type nova 
by spectroscopic observations carried out during the maximum and early 
decline phase. 
\cite{2016MNRAS.455L..57M} presented observations of the time evolution 
in line profiles during the pre-maximum phase, 
and pointed out the presence of pre-existing circumstellar
materials around the nova.
They concluded that the companion in the nova system 
 is not a main sequence dwarf as in most novae but an evolved sub-giant.

In this paper, we report the detection of strong \bevii\ absorption lines in these two novae.
In Section 2, we describe our high resolution spectroscopic observations and data analysis.
The results from our observations are described in Section 3.
The \bevii\ abundances estimated from our observed data are 
presented in Section 4.
A brief discussion and conclusions are given in Section 5.

\section{Observations}

\begin{figure}
\centering
\includegraphics[width=0.95\columnwidth]{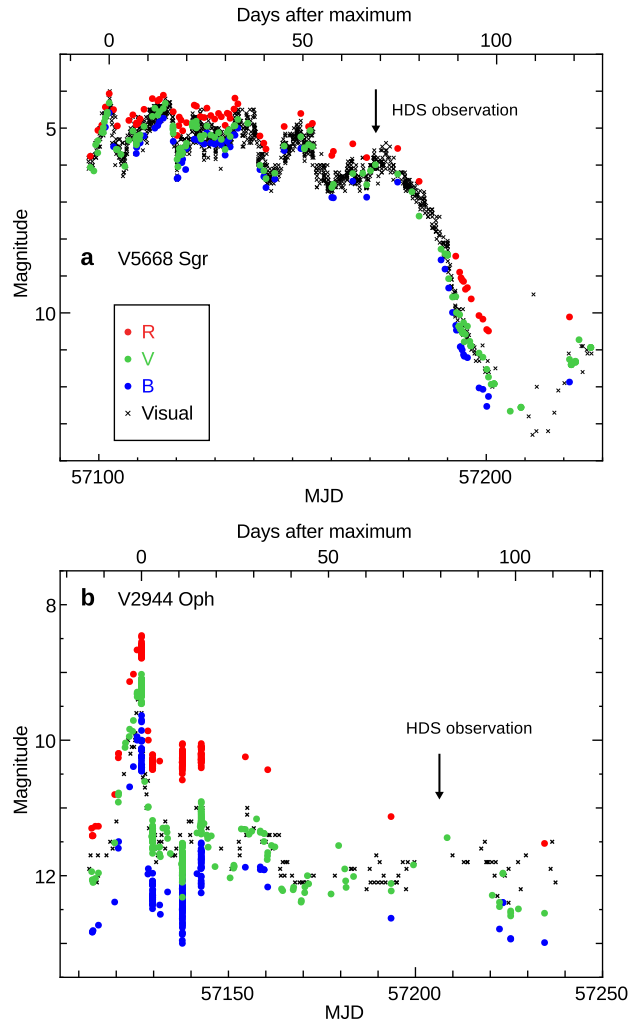}
\caption{The light curves of V5668 Sgr (a) and V2944 Oph (b) observed in Visual, $B$, $V$, and $R$ bands by the AAVSO team. The different bands are plotted in different colors and symbols (see legends in the figure).
The epochs of our HDS observations are indicated by arrows.
} \label{fig-lightvurve}
\end{figure}

The post-outburst spectra of the two classical novae were 
obtained using the High Dispersion Spectrograph (HDS) (\citealt{2002PASJ...54..855N}) of the 8.2-m Subaru Telescope on 2015 May 29 (V5668 Sgr; +69 d after the optical maximum) and July 3 (V2944 Oph; +80 d), respectively.
Figure\,1 shows the AAVSO light curves of two novae with the epochs 
of our HDS observations.
In the case of V5668 Sgr, our observation was just before
 the start of the rapid decline in
optical magnitudes by dust formation.
For each nova, we obtained spectra under 3 configurations of the spectrograph,
which cover the wavelength regions from 3030 to 4630 \AA,
from 4110 to  6860 \AA, and from 6670 to 9360 \AA.
Spectral resolving power was set to $R \simeq\ 60000$ and 45000 with
0\arcsec.6 (0.3 mm) and 0\arcsec.8 (0.4 mm) slit widths,  
respectively.
The details of observations are reported in Table 1.
Data reduction was carried out using the IRAF software in a standard manner.
The non-linearity in the detectors are corrected by 
the method described in \cite{2010PNAOJ...13..1}.
The typical residual of wavelength calibration performed using Th-Ar
comparison spectrum is smaller than $10^{-3}$ \AA\ for each spectrograph 
configuration.
Spectrophotometric calibration was performed using the spectrum of
$\sigma$ Sgr ($V = 2.058$, B2V)
obtained nearly at the same altitude of the nova
on the same nights.
All spectra were converted to the heliocentric scale.
Correction for interstellar extinction has not been applied.
The line identifications were carried out using the atomic line database 
of \cite{NIST_ASD}.
We also used the line list of \cite{1995KurCD..23.....K} for some
weak transitions of Fe-peak elements.

\begin{table}[t]
\caption{Journal of HDS observations of two novae \label{table-obs}}
\begin{center}
\scriptsize
\begin{tabular}{lrrrlr}
\hline
\hline
\multicolumn{1}{c}{Date} & 
\multicolumn{1}{c}{{UT}\tablenotemark{a}} & 
\multicolumn{1}{c}{MJD} & 
\multicolumn{1}{c}{Exposure}& 
\multicolumn{1}{c}{Range}&
\multicolumn{1}{c}{Resolution}\\
\multicolumn{1}{c}{2015}& 
\multicolumn{1}{c}{(h m)} & 
 & 
\multicolumn{1}{c}{(s)} & 
\multicolumn{1}{c}{(\AA)} & \\
\hline
\multicolumn{6}{c}{\bf V5668 Sgr}\\
\hline
May 29            & 10 43 & 57,171.447 &    600 & 4110-6860 & 60,000\\
(+69 d)\tablenotemark{b} & 12 03 & 57,171.502 &  3,600 & 3030-4630 & 60,000\\
                  & 13 57 & 57,171.581 &    300 & 6670-9360 & 60,000\\
\hline
\multicolumn{6}{c}{\bf V2944 Oph}\\
\hline
Jul 03             & 6 55  & 57,206.289 &     900 & 6670-9360 & 45,000\\
(+80 d)\tablenotemark{c}  & 7 10  & 57,206.299 &     900 & 4110-6860 & 45,000\\
                   & 7 44  & 57,206.322 &   5,400 & 3030-4630 & 45,000\\
\hline
\end{tabular}
\tablenotetext{1}{UT is the universal time at the start of an exposure.}
\tablenotetext{2}{Days after the optical ($V$) maximum (MJD = 57,102.67).}
\tablenotetext{3}{Days after the optical ($V$) maximum (MJD = 57,126.75).}

\end{center}

\end{table}





\section{Results}
\begin{figure}
\centering
\includegraphics[width=1.0\columnwidth]{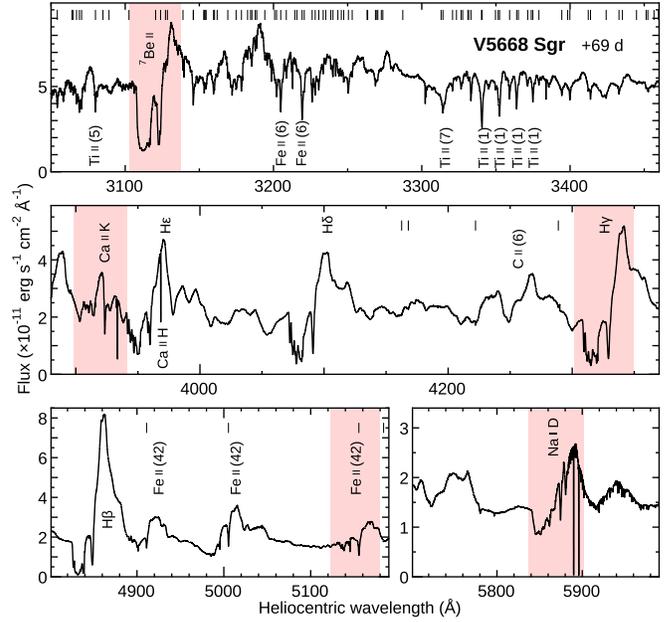}
\caption{Example sections of the flux calibrated HDS spectrum of 
V5668 Sgr obtained at +69 d.
The enlarged views of the spectrum in the hatched areas are presented
in Figure\,\ref{fig-novasgr}. 
Identified absorption line systems originating from Fe-peak elements 
are indicated by ticks at the top.
} \label{fig-novasgr_wide}
\end{figure}

\begin{figure}
\centering
\includegraphics[width=1.0\columnwidth]{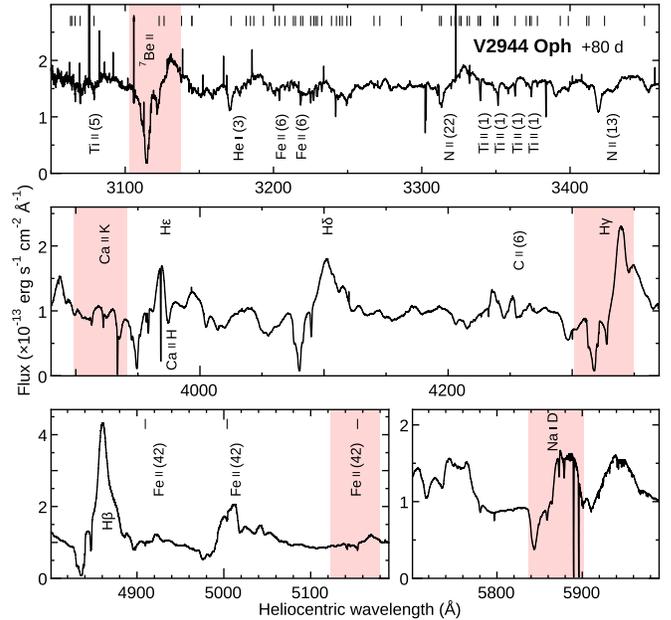}
\caption{Example sections of the flux calibrated HDS spectrum of V2944 Oph 
obtained at +80 d.
The enlarged views of the spectrum in the hatched areas are presented
in Figure\,\ref{fig-novaoph}. 
} \label{fig-novaoph_wide}
\end{figure}

The post-outburst spectra of V5668 Sgr (+69 d)
and V2944 Oph (+80 d), which are displayed in Figure\,\ref{fig-novasgr_wide} and Figure\,\ref{fig-novaoph_wide}, respectively,
 show very similar spectral characteristics.
They contain a series of strong broad emission lines originating from
 \hi\ Balmer series and some other permitted transitions originating from \feii.
As in the case of V339 Del described in \cite{2015Natur.518..381T},
 most of these broad emission lines are accompanied
by blue-shifted multiple absorption lines, whose radial velocities 
correspond to $\sim -2000$ -- $-800$ \kms.
Many blue-shifted absorption lines of Ti\,{\sc ii}, Cr\,{\sc ii}, and Fe\,{\sc ii} are found in the near UV region.

\begin{figure}
\centering
\includegraphics[width=0.85\columnwidth]{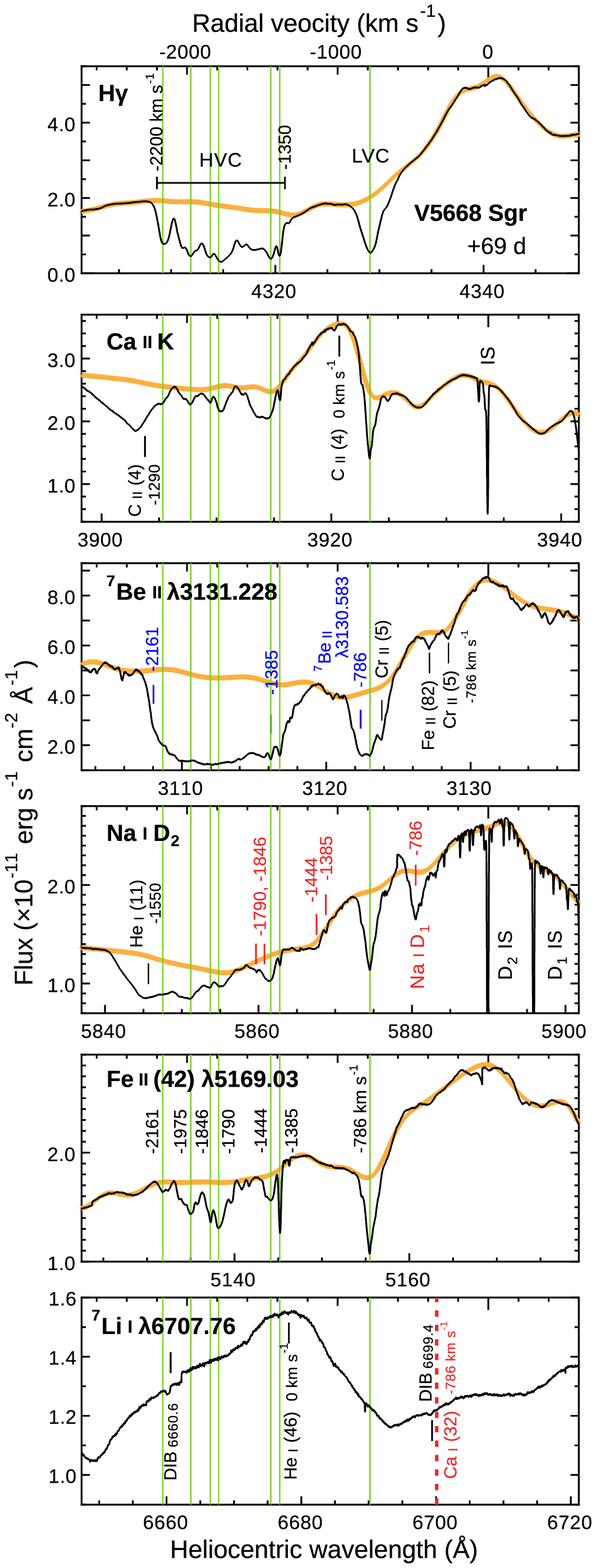}
\caption{The blue-shifted absorption line systems in the spectrum of V5668 Sgr.
The spectrum obtained at +69 d is displayed in the vicinities of \hg, \caii\,K, \beviiii\,$\lambda$3131.228, \nai\,D$_{2}$, \feii\,(42)\,$\lambda$5169.03, and \liviii\,$\lambda$6707.76 lines (top to bottom), on the velocity scale (upper horizontal).
The local continua, fitted with high-order (15--25) spline functions, are
over-plotted with thick lines (orange in the online version).
Dips of individual absorption line, which can be identified in \feii\ and/or \hg, are indicated with vertical lines (green in the online version).
The identified dips by the other component of the doublet are also indicated in the \beviiii\ and \nai\,D panels.
In the bottom panel, no counterparts of the blue-shifted absorption line systems 
of the \liviii\ or the Ca\,{\sc i}\,(32)\,$\lambda$6717.69 lines are found
 in their expected wavelengths (solid and dashed vertical lines).
} \label{fig-novasgr}
\end{figure}

\begin{figure}
\centering
\includegraphics[width=0.85\columnwidth]{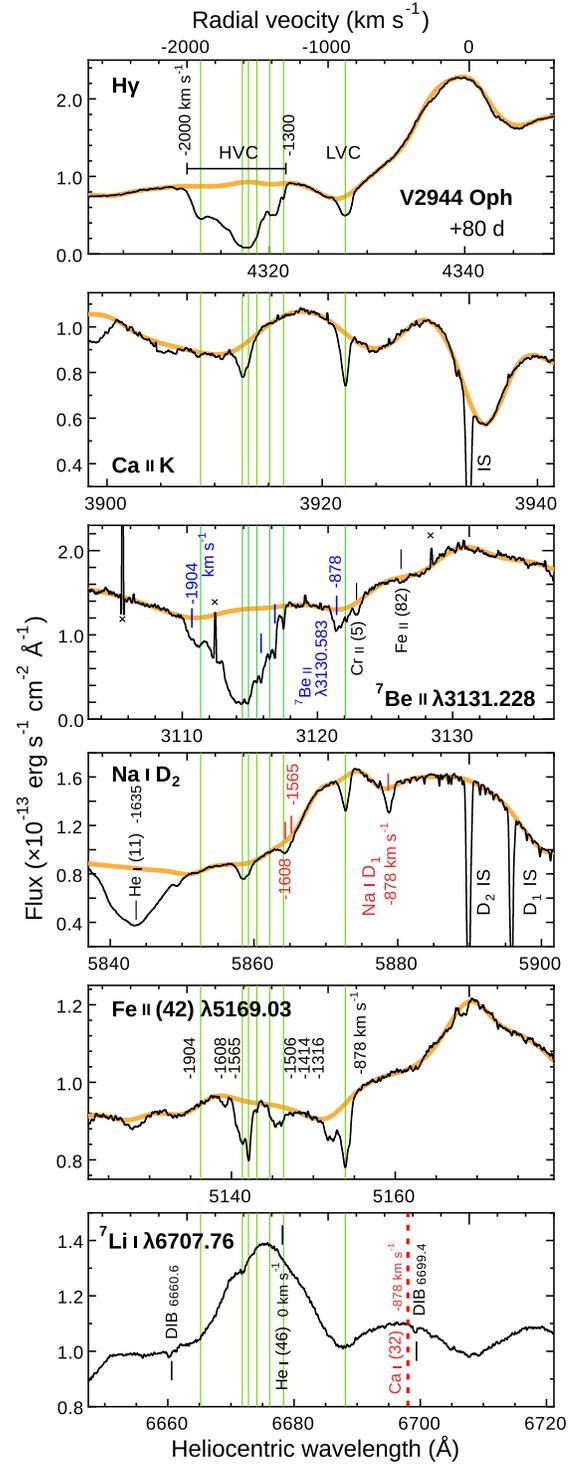}
\caption{Same as Figure\,\ref{fig-novasgr}, but for the spectrum of V2944 Oph obtained at +80 d.
The spectrum is smoothed by a 3 pixel boxcar.
Cosmic-ray hits on the spectrum in the vicinity of \beviiii\ are indicated with crosses.
In the bottom panel, there are no counterparts of the \liviii\ doublet 
or the Ca\,{\sc i}\,(32) lines.
} \label{fig-novaoph}
\end{figure}

\begin{figure*}[t]
\centering
\includegraphics[width=1.2\columnwidth]{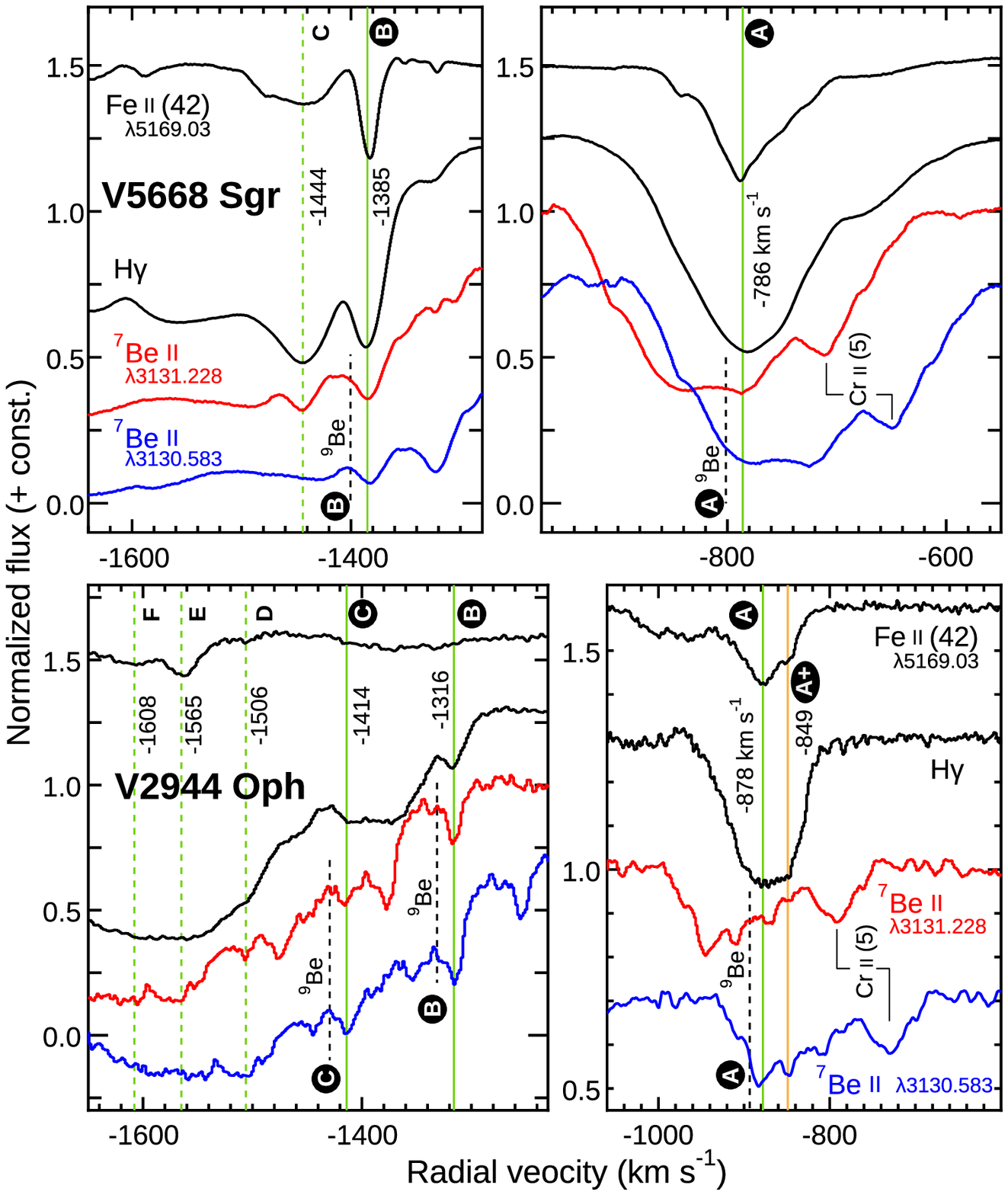}
\caption{Enlarged radial velocity profiles of the LVC and the red side edge of the HVC in V5668 Sgr (upper) and V2944 Oph (lower).
Normalized spectra in the vicinities of \feii\,(42)\,$\lambda$5169, H$\gamma$, and the \beviiii\ doublet are displayed with offsets.
The solid vertical lines indicate absorption components identified in the \beviiii\ doublet and other lines.
The dashed vertical lines show absorption components identified only in \feii\ and/or Balmer lines.
The expected positions of the $^{9}$Be\,{\sc ii} doublet are indicated by partial dashed lines.
In the LVC in V2944 Oph (lower-bottom panel), the spectrum in the vicinity of \beviiii\ lines is smoothed by a 3 pixel boxcar, and 
a sub-component at $v_{\rm rad} = -849$ \kms\ is found in each line (the dip A+; the orange vertical line in the online version).
} \label{fig-dip}
\end{figure*}

Figure\,\ref{fig-novasgr} displays the enlarged views of the spectrum of 
V5668 Sgr in the vicinities of the \hg, \caii\,K, \beviiii\,$\lambda$3131.228,
\nai\,D$_{2}$, \feii\,(the multiplet number 42; \citealt{1959mtai.book.....M})\,$\lambda$5169.03, and Li\,{\sc i} lines, on the velocity scale.
The blue-shifted absorption lines of the \hg\ line is clearly found to be 
divided into two components; the low velocity component (LVC) is a single 
sharp absorption line at a relatively low \vrad $\sim -786$ \kms, 
and the high velocity component (HVC) contains a series of sub-components 
spreading between $-2200 <$ \vrad $< -1350$ \kms.
The radial velocities of each sub-component are measured at the absorption dips in the \feii\,(42) and \hg\ line,
and indicated with vertical lines in the figure.
The sharp absorption of the LVC can be easily identified in all lines 
in the figure.
These LVC lines originating from singly ionized Fe-peak 
elements (\feii, Ti\,{\sc ii}, Cr\,{\sc ii}, Mn\,{\sc ii}, and Ni\,{\sc ii}) 
dominate the spectrum in the UV range as in the case of V339 
Del from +38 to +48 d.
All of the identified transitions originate from levels of
low excitation potentials (the lower energy level of the transition, $E_{\rm lower} \lesssim$ 3.1 eV).
The absorption strengths of the HVC vary among different transitions.
All of six sub-components in the HVC
can be identified in the \hg\ and \feii\,(42) lines, 
although some of them are weaker or missing in \caii\ and \nai\ lines.
In the UV range, only the sub-components at \vrad $= -1385$ and $-1444$ \kms\ 
of the HVC can be identified in the strong lines originating from Fe-peak
elements. 

Figure\,\ref{fig-novaoph} presents the spectrum of V2944 Oph 
 displayed in the same way as in Figure.\,\ref{fig-novasgr}.
The spectrum of this nova shows quite similar characteristics to 
that of V5668 Sgr.
It also shows the blue-shifted absorption systems divided 
into two velocity components; the LVC at 
\vrad $= -878$ \kms\ and the HVC between $-2000 <$ \vrad\ $< -1300$ \kms.
At least six sub-components can be identified in the HVC 
as shown by vertical lines in the figure.

As shown in the bottom panels of Figure \ref{fig-novasgr} and \ref{fig-novaoph},
 we cannot find any counterparts of blue-shifted absorption line systems of
 the \liviii\,$\lambda$6708 in spite of the high signal-to-noise ratios 
 of the spectra ($\sim700$ and $\sim160$ at 6690 \AA\ in V5668 Sgr and 
 V2944 Oph, respectively).
There is also no counterpart of the Ca\,{\sc i}\,(32)\,$\lambda$6718
 in the spectra of both novae, although this line has been detected 
 together with the \lii\,$\lambda$6708 line in the first three weeks spectra 
 of V1369 Cen  (\citealt{2015arXiv150608048I}).
The \cai\,$\lambda$4227 and K\,{\sc i}\,$\lambda$7699 lines
 do not have any counterparts in the LVC and the HVC in both observed novae.
These observations imply that the ejected gas had been heated into so hot
 state that almost all Li and Ca atoms are ionized in the circumstance
 during epochs of our HDS observations.

Figure\,\ref{fig-dip}  displays 
the enlarged radial velocity profiles of
\feii, \hg, and the \beviiii\ ``doublet'' lines in both novae.
Although the \beviiii\,$\lambda\lambda$3130.583,\,3131.228 doublet lines 
are coalesced each other, 
the dips of each line can be partially identified.
In V5668 Sgr, two absorption dips (A and B) are clearly found in all lines.
In V2944 Oph, two dips (B and C) of the \beviiii\ doublet coincide with 
those of \hg. 
The dip A of \beviiii\,$\lambda$3131.228 (the weaker component of the doublet)
 in the LVC in this nova is unclear 
 due to the poor signal-to-noise ratio, which is partly disturbed by
 cosmic ray hits on the spectrum. 
A sub-component in the LVC at $v_{\rm rad} = -849$ \kms\ (the dip A+) is found in
the \feii\ and both \beviiii\ lines.
We can find no alternative candidates of identifications for
 these dips from line lists of Fe-peak elements.
Because no dips corresponding to the $^{9}$Be\,{\sc ii} doublet [the isotopic shift; $\Delta\lambda = -0.161$ \AA\ (\citealt{2008PhRvL.100x3002Y})] can be found,
we conclude that the absorption systems at $\sim$ 3110 -- 3125 \AA\ 
in both novae are purely originating from \beviiii\ as in the case of V339 Del.
The dips of the \beviiii\ doublet become hard to be identified 
toward the flat bottoms of the HVC in both novae.
This indicates that the absorption of \beviiii\ is saturated,
while the intensities at the flat bottoms remain $\sim$25 and $\sim$15 \% 
of the continua in V5668 Sgr and V2944 Oph, respectively.
In V5668 Sgr, the HVCs of the Balmer lines show wavy bottom shapes.
This difference indicates that 
the \beviiii\ absorption is stronger 
than the Balmer lines at this velocity range.

It is interesting to see the similarity in the velocity profiles of 
the blue-shifted absorption systems in both novae.
Some of classical novae also have similar blue-shifted absorption systems
 with two distinct velocity components -- the narrow LVC and the broad HVC 
(e.g., Nova LMC 2005, V2574 Oph in \citealt{2008ApJ...685..451W}), 
which are known as the ``principal'' and the ``diffuse enhanced'' 
systems in post-maximum nova spectra (\citealt{1960stat.conf..585M,2014ASPC..490..183M}).
V1369 Cen at +13 d also has two velocity components at \vrad\ $=-550$ and
$-1300 <$ \vrad\ $<-900$ (\citealt{2015arXiv150608048I}).
One remarkable point found in this study is 
that \bevii\ is detected in both the LVC and the HVC.
This means that both of the absorption systems consist of nova ejecta
 which have experienced TNR.
The characteristics of the LVCs in V5668 Sgr and V2944 Oph 
closely coincide with those of the Transient
 Heavy Element Absorption (THEA) found by 
\cite{2008ApJ...685..451W} in post-outburst spectra of classical novae.
They assumed that these THEAs are produced by pre-outburst ejecta 
coming from the secondary star.
Considering our results on the \bevii\ detection, however,
it is natural to assume that the THEAs are produced by nova ejecta 
which have experienced TNR.

\begin{table*}[t]
\begin{center}
\caption{Lines originating from Fe-peak elements in the vicinity of the \beviiii\ doublet \label{table-contami}}
\begin{tabular}{lcrccccccc}
\hline
\hline
& & & \multicolumn{3}{c}{V5668 Sgr} & &\multicolumn{3}{c}{V2944 Oph} \\[2pt] \cline{4-6} \cline{8-10} 
\multicolumn{1}{c}{Lines}&
\multicolumn{1}{c}{$\lambda_{\rm lab}$}&
\multicolumn{1}{c}{$\log{(gf)}$}&
\multicolumn{1}{c}{$\lambda_{\rm LVC}$\tablenotemark{a}}& 
\multicolumn{1}{c}{\beviiii\tablenotemark{b}}&
\multicolumn{1}{c}{$W$ (\AA)\tablenotemark{c}}&
&
\multicolumn{1}{c}{$\lambda_{\rm LVC}$\tablenotemark{a}}& 
\multicolumn{1}{c}{\beviiii\tablenotemark{b}}&
\multicolumn{1}{c}{$W$ (\AA)\tablenotemark{c}}\\[2pt] \hline
Ti\,{\sc ii}\,(67) & 3106.246 & $-0.170$ &  3098.11  & --- &  $<$ 0.003           & &  3097.16  & --- &  $<$ 0.006          \\
Ti\,{\sc ii}\,(67) & 3117.676 & $-0.500$ & (3109.51) & HVC &  ---                 & & (3109.51) & HVC &  ---                \\
Cr\,{\sc ii}\,(5)  & 3118.646 & $ 0.000$ & (3110.48) & HVC &  (0.054 $\pm$ 0.026) & & (3109.52) & HVC &  (0.037 $\pm$ 0.020)\\
Ti\,{\sc ii}\,(67) & 3119.825 & $-0.460$ & (3111.65) & HVC &  ---                 & & (3110.69) & HVC &  ---                \\
Cr\,{\sc ii}\,(5)  & 3120.359 & $+0.120$ & (3112.18) & HVC &  (0.071 $\pm$ 0.034) & & (3111.23) & HVC &  (0.049 $\pm$ 0.030)\\
Cr\,{\sc ii}\,(5)  & 3120.497 & $-0.018$ & (3112.32) & HVC &  (0.052 $\pm$ 0.025) & & (3111.36) & HVC &  (0.036 $\pm$ 0.020)\\
Cr\,{\sc ii}\,(5)  & 3128.692 & $-0.320$ & 3120.49   & --- & 0.013 $\pm$ 0.003    & & 3120.49   & --- & $<$ 0.006   \\
Cr\,{\sc ii}\,(5)  & 3132.053 & $+0.079$ & 3123.85   & LVC & 0.067 $\pm$ 0.015    & & 3123.85   & --- & 0.045 $\pm$ 0.015   \\
Cr\,{\sc ii}\,(5)  & 3136.681 & $-0.250$ & 3128.46   & --- & 0.045 $\pm$ 0.015    & & 3128.46   & --- & $<$ 0.006   \\
\hline
\end{tabular}
\tablenotetext{1}{The observed and expected wavelengths of the LVCs (\vrad\ $= -786$ and $-878$ \kms in V5668 Sgr and V2944 Oph, respectively) of each transition. The undetected lines in the \beviiii\ HVC are listed in brackets.}
\tablenotetext{2}{The velocity component of the \beviiii\ doublet expected to be contaminated by the line.}
\tablenotetext{3}{The observed or expected equivalent widths of each line. 
The values estimated using nearby same multiplets are listed in brackets.}
\end{center}

\end{table*}

The \beviiii\ absorption may be contaminated by LVCs originating from 
other Fe-peak elements.
We set requirements for major contaminants as 
$E_{\rm lower} < 3.1$ eV and $\log{(gf)} > -1.0$, 
and select candidates for contaminants as listed 
in Table\,\ref{table-contami}.
Only some Cr\,{\sc ii}\,(5) lines are found to be the main
contaminants to the LVC and the HVC of the \beviiii\ doublet.
In V5668 Sgr,
the dip of the Cr\,{\sc ii}\,(5)\,$\lambda$3132.053 is clearly 
identified within the \beviiii\ LVC (Fig.\,\ref{fig-dip} upper-right panel).
In the \beviiii\ HVC, no Cr\,{\sc ii}\,(5) lines can be identified because of
the strong saturation effect.
However, their equivalent widths ($W$) 
 can be estimated using measured equivalent widths of nearby unblended 
 Cr\,{\sc ii} lines belonging to the same multiplets.
We estimate that the combined contributions of these contaminants 
are less than 5\%\
of the total equivalent widths of the \beviiii\ absorption 
in both the LVC and the HVC in V5668 Sgr.
In the case of V2944 Oph, we find that only 
the Cr\,{\sc ii}\,(5)\,$\lambda$3132.053 is present besides the LVC
 of the \beviiii\ doublet.
We estimate that the combined contribution of contaminants 
in the HVC in this nova is similar to or less than that in V5668 Sgr.

\begin{table*}[t]
\begin{center}
\caption{Equivalent widths of blue-shifted absorption lines in observed novae \label{table-ew}}
\begin{tabular}{lrrccccc}
\hline
\hline
\multicolumn{1}{c}{}&
\multicolumn{1}{c}{}&
\multicolumn{1}{c}{HVC} & 
\multicolumn{1}{c}{}&
\multicolumn{4}{c}{LVC}\\ \cline{3-3} \cline{5-8}
\multicolumn{1}{c}{\vspace{-0.3cm}Object} & 
\multicolumn{1}{c}{Day \vspace{+0.3cm}} & 
\multicolumn{1}{c}{$\begin{array}{c}W(\mbox{\beviiii})\\(\mbox{\AA})\end{array}$}& 
\multicolumn{1}{c}{}&
\multicolumn{1}{c}{$\begin{array}{c}W(\mbox{\beviiii})\\(\mbox{\AA})\end{array}$}& 
\multicolumn{1}{c}{$\begin{array}{c}W(\mbox{\caii\,K})\\(\mbox{\AA})\end{array}$}& 
\multicolumn{1}{c}{$\begin{array}{c}N(\mbox{\beviiii})\\\hline N(\mbox{\caii})\end{array}$}& 
\multicolumn{1}{c}{$\begin{array}{c}X(\mbox{\bevii})\\\hline X(\mbox{Ca})\end{array}$}\\ 
\hline
V5668 Sgr & +69 d & $6.40 \pm 0.98$ && $1.30 \pm 0.22$ & $0.35 \pm 0.02$ & $8.1 \pm 2.0$ & $1.4 \pm 0.30$\\
V2944 Oph & +80 d & $3.64 \pm 0.30$ && $0.20 \pm 0.10$ & $0.18 \pm 0.03$ & $2.5 \pm 1.5$ & $0.44 \pm 0.26$\\
V339 Del & +47 d\\
\multicolumn{2}{c}{\small $-$1103 \kms}  &  && $0.16 \pm 0.010$ & $0.089 \pm 0.003$ & $3.9 \pm 0.4$ & $0.68 \pm 0.07$\\
\multicolumn{2}{c}{\small $-$1268 \kms}  &  && $0.064 \pm 0.015$ & $0.023 \pm 0.004$ & $6.5 \pm 2.5$ & $1.1 \pm 0.44$\\
\hline
\end{tabular}
\end{center}
\end{table*}

\section{\bevii\ abundance}
We placed ``best-effort'' local continua 
around each line ($ -3000 < v_{\rm rad} < 1000$ \kms) on the complex,
undulating spectra of two novae
 by the fitting with a high-order (15 -- 25) spline
function to measure the equivalent widths of the absorption lines.
Although the \beviiii\ doublet cannot be resolved, 
its total equivalent widths, $W(\mbox{\beviiii})$s, 
in the LVC and the HVC are measured.
After subtracting the combined effects of contaminations 
discussed above,
we tabulate the resulting $W(\mbox{\beviiii})$ in Table\,\ref{table-ew}.
The errors of $W$ are estimated from the signal-to-noise ratios of 
the neighboring-continuum.
They do not include the uncertainties in the continuum placements.

The abundances of \bevii\ in nova ejecta is estimated 
following the same procedure as used in \cite{2015Natur.518..381T}.
We compare the equivalent widths of \caii\,K ($\log{(gf)} = +0.135$) and \beviiii\,$\lambda3130.583+\lambda3131.228$ ($\log{(gf)} = -0.178, -0.479$, 
respectively) in the LVC, which show a simpler structure than the HVC.
Here, we need the following assumptions;  (1) both the \beviiii\ and \caii\ lines are not saturated,
(2) the covering factor of the absorbing gas cloud to the
background illuminating source has no wavelength dependence.
We notice that the LVC feature of \beviiii\ at $-786$ \kms\ of V5668 Sgr
might be saturated.
The bottom of the \beviiii\ lines in the LVC almost reaches to
that in the HVC, where its flat profile suggests that the absorption
is saturated.
The corresponding LVC component of the \caii\,K line in V5668 Sgr is shallow
and the line appears to be unsaturated.
The abundance of \bevii\ in V5668 Sgr obtained under such a condition 
should be taken as a lower limit.

Following \citeauthor{1998ppim.book.....S} (\citeyear{1998ppim.book.....S}, Equation (3)-(48)),
the ratio of column number densities, $N$, can be written as
\begin{eqnarray}
  \frac{N(\mbox{\beviiii})}{N(\mbox{\caii})}
  & = &\frac{W(\mbox{\beviiii\,})}{\lambda(\mbox{\beviiii})^{2}} 
  \times \frac{\lambda(\mbox{\caii\,K})^{2}}{W(\mbox{\caii\,K})} \times \nonumber\\  
  &   & \, \frac{gf(\mbox{\caii\,K})}{gf(\mbox{\beviiii\,}\lambda3130.583) + gf(\mbox{\beviiii\,}\lambda3131.228)} \nonumber\\
  & = & \frac{W(\mbox{\beviiii\,})}{3131^{2}}
  \times \frac{3934^{2}}{W(\mbox{\caii\,K})} \nonumber\\
  &   & \, \times \frac{10^{+0.135}}{10^{-0.178}+10^{-0.479}} .
\end{eqnarray}
Using data of the $W(\mbox{\beviiii})$ and the $W(\mbox{\caii\,K})$
 listed in Table\,\ref{table-ew},
the column density ratios,  $N(\mbox{\beviiii})/N(\mbox{\caii})$, 
derived in the LVC of V5668 Sgr and V2944 Oph are 
  $8.1 \pm 2.0$  and  $2.5 \pm 1.5$, respectively.
These ratios are close to that found in V339 Del at +47 d.

The ionization states of each species are keys to convert the above ratios
 into atomic abundances.
We assume that the most of \bevii\ is in the singly ionized state,
because it still stays in hot ejecta.
We also assume that Ca is in the singly ionized state.
The absence of \cai\ lines in the blue-shifted absorption line systems
 supports this assumption.
The difference in the second ionization potential (Be: 18.21 eV, Ca: 11.87 eV)
  may lead to different balances between the second and the third ionized 
 ions and this may be a source of an error in determining the abundances.
We tried but failed to detect absorption 
lines of doubly ionized ions of Fe-peak elements, 
which have intermediate second ionization potentials,
consulting \cite{NIST_ASD}.
Absorption lines of Fe\,{\sc iii} are 
observed in optical range spectra of B-type stars 
(\citealt{2008MNRAS.383..729T}).
This may support the assumption that all of \bevii\ and Ca contained 
in the LVC are in the singly ionized state. 
Adopting this assumption, 
the mass fraction of \bevii\ to the sum of all constituent mass components, 
$X($\beviiii$)$, can be presented as: 
$X($\bevii$) \sim N($\beviiii$)/N($\caii$) \times 7/40 \times X($Ca$)$.
We have obtained $X($\bevii$)/X($Ca$) = 1.4 \pm 0.3$ for V5668 Sgr
and $0.44 \pm 0.26$ for V2944 Oph.
These $X(\mbox{\bevii})/X(\mbox{Ca})$ values correspond to 
logarithmic overabundances of Li by +6.2 dex and +5.7 dex
with respect to the solar photospheric abundance 
(\citealt{2009ARAaA..47..481A}). 
As in the case of V339 Del, 
the Li abundances in ejecta of both nova explosions are comparable 
to or even higher than those of Ca.
Because the results presented here are based on the observations 
$\sim$ 75 -- 100 days after 
their outbursts, the amount of \bevii\ freshly produced in the TNR could 
be 3 -- 4 times larger than our measurements.

We note that these abundances are obtained only using a limited fraction of 
the nova ejecta giving rise to the LVC absorption.
They do not necessarily represent the abundances in the whole materials erupted
from these novae.
Furthermore, when the \beviiii\ lines are saturated, 
the above $X(\mbox{\bevii})/X(\mbox{Ca})$ values should be treated carefully.
Therefore, the derived \bevii\ abundances in this study 
might involve large uncertainties.
However, the above result implies that classical novae are indeed playing 
an important role in the process of the Galactic Li enrichment.

To produce \bevii\ via the reaction \iiihe($\alpha$, $\gamma$)\bevii,
 the total number of \iiihe\ atoms in the accreted gas before TNR 
 should be larger than or equal to that of freshly produced \bevii\ atoms
 in nova ejecta.
This means that the number ratios, $N(\mbox{\iiihe})/N(\mbox{Ca})$, in the 
accreted gases of V5668 Sgr and V2944 Oph are $\gtrsim 8.1 \pm 2.0$ and 
 $\gtrsim 2.5 \pm 1.5$, 
 respectively.
If we adopt the solar abundance ratio between He and Ca
 ($\log{\left( N(\mbox{He})/N(\mbox{Ca}) \right) } = +4.59$; 
  \citealt{2009ARAaA..47..481A}),
 the lower limits of 
 the number ratios between \iiihe\ and \ivhe, $N(\mbox{\iiihe})/N(\mbox{\ivhe})$,  
 in the accreted gases of V5668 Sgr and V2944 Oph should amount to $0.020\pm0.005$ and $0.006\pm0.003$\%, respectively.
These isotopic ratios are close to that found in the solar system 
(= 0.0166\%;  \citealt{2009ARAaA..47..481A}).
\cite{1996ApJ...465L..27H} pointed out that \iiihe\ destruction
 via the reaction \iiihe(\iiihe, 2$p$)\ivhe\ occurs concurrently with 
\iiihe($\alpha$, $\gamma$)\bevii\ during TNR.
They noted that the \iiihe\ destruction via 
 \iiihe(\iiihe, 2$p$)\ivhe\ is less pronounced in CO novae 
 than in ONe novae.
Because both V5668 Sgr and V2944 Oph are found to be CO novae (see the
next section), we guess that the process of \iiihe\ destruction
 should be ineffective and that most of \iiihe\ in the accreted gases 
 have been converted into \bevii\ in both novae.

\section{Discussion and conclusions}

The detection of \bevii\ in the post-outburst spectra of 
 V5668 Sgr and V2944 Oph strongly suggests that 
 the explosive production of \bevii\ is a popular phenomenon among classical novae.
The light curves of both novae show slower evolution than 
that of V339 Del in which the rapid decline 
were observed in $\sim$ +40 d.
Adopting the relation between the time-scaling factor of the light curve and
the WD mass (e.g., \citealt{2006ApJS..167...59H,2015ApJ...798...76H}),
the WD masses of both novae are lower than that of V339 Del.
It is natural to assume that both novae have CO WDs 
as in the case of V339 Del.
Therefore, we infer that the explosive Li production via the Cameron-Fowler
 process occurs frequently at least among CO novae.

It is interesting to notice that we have found LVC and HVC absorptions
of the \nai\,D lines in V5668 Sgr and V2944 Oph, 
while no trace of blue-shifted absorptions of the \nai\,D lines 
were detected in V339 Del (\citealt{2015Natur.518..381T}).
We find no absorption features corresponding to 
\lii\,$\lambda$6708, K\,{\sc i}\,$\lambda$7699, and \cai\,$\lambda$4227
 in both V5668 Sgr and V2944 Oph 
as well as in V339 Del.
\cite{2015arXiv150608048I}
 found blue-shifted absorptions of H$\beta$ and \nai\,D
 lines in V1369 Cen at two velocities ($-1100$ and $-550$ \kms) on  
 spectra obtained during early phase after the outburst (7 to 13 days).
They identified \lii\,$\lambda$6708, K\,{\sc i}\,$\lambda$7699, \cai\,$\lambda$6718 and \cai\,$\lambda$4227 in the lower velocity component.
These apparent spectral differences, 
such as detection or no-detection of absorptions 
of the \nai\,D or \cai\ lines, should reflect differences 
in physical conditions 
(most probably the ionization state) in the absorbing clouds.
On the other hand, the presence of \bevii\ both in the LVC and the HVC
strongly indicates that the absorbing clouds in both velocity 
components consist of materials which have experienced TNR.

The abundance of Li obtained by \cite{2015arXiv150608048I}
 from the \lii\,$\lambda$6708 is lower than the \bevii\ (= Li) 
 abundances obtained in the present study.
There had been no report in literature 
of detection of the \lii\,$\lambda$6708 line among post-outburst novae 
before \cite{2015arXiv150608048I}. 
Because the \lii\,$\lambda$6708 is located in the easily accessible 
red spectral region, the paucity of observational information 
may imply that ejecta of post-outburst novae rarely permits
the physical condition in which neutral Li can survive. 
However, it will be interesting to carry out long term monitoring 
observations of post-outburst spectra of novae including both \bevii\ and 
Li lines and examine the consistency of derived Li abundances. 
Further observations of the \beviiii\ line in various 
types of classical novae will help to quantify 
roles of novae as the source of Li in the current universe.



\acknowledgments

This work is based on data observed at the Subaru Telescope,
which is operated by the National Astronomical Observatory of Japan
(NAOJ).
We acknowledge with thanks the variable star observations from the AAVSO
International Database contributed by observers worldwide and used in
this research. 



{\it Facilities:} \facility{Subaru(HDS)}.

\end{document}